\pdfoutput=1

\documentclass[aps,prb,reprint,showpacs,superscriptaddress]{revtex4-1}
\usepackage{graphicx}
\usepackage{amsmath}
\usepackage{amssymb}
\usepackage{amsfonts}
\usepackage{dcolumn}
\usepackage{dsfont}
\usepackage{latexsym}
\usepackage{rotating}
\usepackage{color}
\usepackage{latexsym}
\usepackage{bbm}
\usepackage{subfigure}
\usepackage{float}
\usepackage{epsfig}
\usepackage{psfrag}
\usepackage{natbib}
\usepackage{bm}
\usepackage{amsthm}
\usepackage{eucal}
\usepackage{mathrsfs}
\usepackage{url}
\usepackage{braket}

\usepackage{color} % use colored text in latex

\usepackage{hyperref}
\hypersetup{
colorlinks=true,final=true,
        linkcolor=blue,
        citecolor=blue,
        filecolor=blue,
        urlcolor=blue,
}
\begin{document}

\title{Anomalous transport near the Lifshitz transition at the {L}a{A}l{O}$_3$/{S}r{T}i{O}$_3$ interface}
\author{S. Nandy}
\affiliation{Department of Physics, Indian Institute of Technology Kharagpur, W.B. 721302, India}
\author{N. Mohanta}
\email{narayan.mohanta@physik.uni-augsburg.de}
\affiliation{Center for Electronic Correlations and Magnetism, Theoretical Physics III, Institute of Physics, University of Augsburg, 86135 Augsburg, Germany}
\author{S. Acharya}
\affiliation{Department of Physics, Indian Institute of Technology Kharagpur, W.B. 721302, India}
\author{A. Taraphder}
\affiliation{Department of Physics, Indian Institute of Technology Kharagpur, W.B. 721302, India}
\affiliation{Centre for Theoretical Studies, Indian Institute of Technology Kharagpur, W.B. 721302, India}

\begin{abstract}

The two-dimensional electron liquid, at the (001) interface between band insulators {L}a{A}l{O}$_3$ and {S}r{T}i{O}$_3$, undergoes Lifshitz transition as the interface 
is doped with carriers. At a critical carrier density, two new orbitals populate at the Fermi level, with a concomitant change in the Fermi surface topology. 
Using dynamical mean-field theory, formulated within a realistic three-orbital model, we study the influence of the Lifshitz transition and local electron correlations on 
the transport properties. We look at the thermal conductivity, optical conductivity, Seebeck coefficient and angle resolved photoemission spectra and find that at a 
critical density, both the thermal and dc conductivities rise sharply to higher values while the Seebeck coefficient shows a cusp. The inter-orbital electron-electron interaction 
transfers spectral weight near the $\Gamma$ point towards lower energy, thereby reducing the critical density. In the presence of external magnetic field, the 
critical density further reduces due to exchange splitting. Beyond a sufficiently large field, multiple cusps appear in the Seebeck coefficient revealing multiple Lifshitz 
transitions.
\end{abstract}
           
%\pacs{73.20.-r,  73.50.-h,  71.18.+y, 31.15.V-}

\maketitle

%=============================================
\section{Introduction}
%=============================================
The metallic interface between perovskite band insulators LaAlO$_3$ (LAO) and SrTiO$_3$ (STO) is a prototypical STO-based heterointerface~\cite{Ohtomo2004,Thiel_Science2006,Huijben_NMat2006} having a wide variety of intriguing properties such as superconductivity (below 200 mK)~\cite{Reyren31082007,GariglioJPCM2009,RichterNature2013,Marel_PRB2014,GariglioJPCM2009}, ferromagnetism (below 200 K)~\cite{Li_Nature2011,Dikin_PRL2011,Dagan_PRL2014}, ferroelectricity~\cite{Tra_AdMa2013} and strong spin-orbit coupling (SOC)~\cite{Fischer_NJP2013,Fete_PRB2012,Fete_PRB2012}. The good tunability of these properties~\cite{Caviglia2008,CavigliaPRL2010,Cen_NMat2008,Balatsky_PRB2012,Bucheli_PRB2014}, with respect to external fields, makes this interface a potential candidate for novel device applications~\cite{Nakagawa2006,Dagotto1076,Rijnders_Nature2008}. The electrons in the quasi two-dimensional electron liquid (q2DEL) at the interface are supplied by an intrinsic electronic transfer process, known as the `polar catastrophe mechanism'~\cite{Ohtomo2004,Nakagawa2006,Satpathy_PRL2008}, and also by the oxygen vacancies near the interface~\cite{Ariando_PRX2013,Li_PRB2011,Zhong_PRB2010}. The superconductivity is mediated by interface phonons and is of conventional in nature~\cite{Boschker_SciRep2015}. There are two possible sources of ferromagnetism: (i) as suggested previously~\cite{Michaeli_PRL2012,Banerjee2013}, the electrons occupying the $d_{xy}$ orbital of Ti ions at the TiO$_2$-terminated interface are localized due to the repulsive electron-electron interaction forming a quarter-filled charge-ordered insulating state and the ferromagnetism is a result of an exchange coupling between the localized moments and the conduction electrons residing below the interface layer; (ii) density function theory (DFT) indicates that the oxygen vacancies near the interface lead to a spin splitting of the electrons in the occupied $t_{2g}$ orbitals at the Fermi level giving rise to ferromagnetic order~\cite{Kelly_PRB2014,Pavlenko_PRB2009,Vanderbilt_PRB2009,Park_PRL2013,Pavlenko_PRB2013}. The ferromagnetism, which is tunable~\cite{Stemmer_PRX2012,Kelly_PRB2014} and is of $d_{xy}$ character~\cite{Lee_Nature2013}, coexists with superconductivity~\cite{Bert_NPhys2011,Li_Nature2011,Dikin_PRL2011,CNayak_PRB2013,Pavlenko_PRB2012_1} in spatially phase segregated regions due to the inhomogeneity at the interface~\cite{mohantaJPCM,NM_VacancyJPCM2014,Ariando2011}. The coexistence of these competing orders gives rise to unconventional phenomenon such as the field-induced transient superconductivity~\cite{2014arXiv1411.3103K} and topological superconductivity due to the presence of the Rashba SOC~\cite{0295-5075-108-6-60001,Scheurer2015}. Transport measurements reveal that a new type of carriers, of higher mobility, appears as a new conducting channel as the gate-voltage is increased~\cite{Shalom_PRL2010,Dagan_PRL2014}. These carriers originate from the $d_{yz}$, $d_{zx}$ orbitals, which are separated from the $d_{xy}$ orbital due to confinement at the interface~\cite{Held_PRB2013,Khalsa_PRB2013}, and are believed to be responsible for participating in superconductivity~\cite{Shalom_PRL2010,PhysRevB.92.174531,Nakamura_JPSJ2013,Caprara_PRB2013}. The emergence of the high mobility carriers, with increasing the Fermi level, triggers a Lifshitz transition~\cite{Imada_JPSJ}, in which the topology of the Fermi surface changes, and results in a distinct phase with large off-diagonal Hall conductivity~\cite{Joshua_PNAS2013,Joshua_NComm2012}. The impact of the Lifshitz transition on the transport properties, especially when the interaction between the electrons is not negligible, is not much explored. Furthermore, the role of the electron correlation in the Lifshitz transition is an interesting aspect unaddressed so far.

In this paper, we adopt a model Hamiltonian with six spin-orbit coupled $t_{2g}$ orbitals, developed using DFT analysis~\cite{Held_PRB2013}, and employ dynamical mean field theory (DMFT) with iterative perturbation theory (IPT) as the impurity solver to investigate the transport properties in the presence of inter-orbital electron-electron interaction. The change in topology in the interacting Fermi surface is investigated. We calculate the thermal conductivity, optical conductivity and Seebeck coefficient
 (thermopower) which reveal anomalous behavior at a critical carrier density. As the carrier density is tuned, the thermal and optical conductivities sharply rise to higher values while the Seebeck coefficient exhibits a cusp. We compute the intensity of the angle resolved photoemission spectroscopy (ARPES) which shows that the spectral weight, near the $\Gamma$ point, is transferred towards lower energy. The electron-electron interaction, therefore, slowly decreases the critical density for the Lifshitz transition. We also study the situation in the presence of external magnetic field and find that the critical density further reduces with increasing field strength. However, beyond a large critical field, the spin degeneracy is removed in all the orbitals and multiple cusps appear in the Seebeck coefficient indicating multiple Lifshitz transitions. Unlike the low-field case, the critical densities for the new Lifshitz transitions increase with increasing field strength. We compare our findings with existing experimental data and discuss the possibility to observe these anomalous phenomena in future experiments.

The rest of the paper is organized as follows. In Sec.~\ref{model}, we introduce the effective Hamiltonian for the interface q2DEL and briefly describe the formulation of the multi-orbital DMFT, employed to solve the inter-orbital electron-electron interaction. In Sec.~\ref{results}, we discuss our numerical results establishing the anomalous features observed in the transport properties and the  multiple Lifshitz transitions in the presence of external magnetic field. We also make comparison of our results with existing experimental data. Finally in Sec.~\ref{summary}, we discuss the experimental aspects of the phenomena observed in our study and summarize our results.
 
%=============================================
\section{Model and method}
\label{model}
\begin{figure*}[htb!]
\begin{center}
\epsfig{file=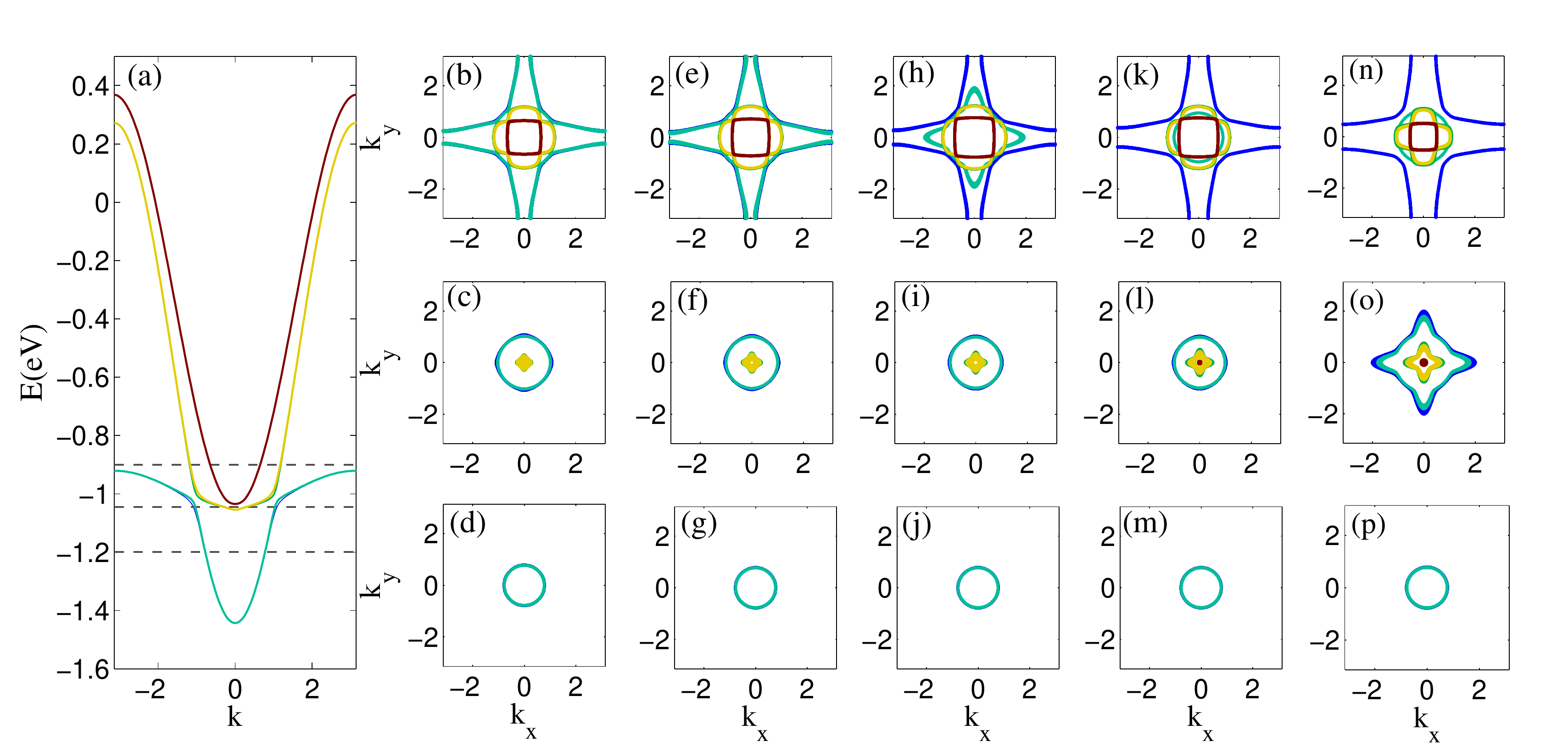,trim=0.00in 0.10in 0.15in 0.1in,clip=false, width=175mm}
\caption{(Color online)(a) The resultant band structure of the q2DEL along $\bar{X}$-$\Gamma$-X (in units of $\pi$) direction at the interface showing the $t_{2g}$ orbitals in presence of the Rashba and atomic SOCs. (d)-(c)-(b) describe the non-interacting ($U=0$) Fermi surfaces below ($n=0.0945$ el./u.c.), near ($n=0.1809$ el./u.c.) and above ($n=0.8030$ el./u.c.) the Lifshitz transition point ($\mu_c=-1.067$ eV) at which the third and fourth bands begin to get occupied. The dashed horizontal lines in (a), at the corresponding three chosen chemical potentials $\mu=-1.20$ eV, $\mu=-1.05$ eV and $\mu=-0.90$ eV respectively, are  guide to the eyes. (e)-(p) show interacting Fermi surfaces at 60K at the interface. Columns (e) to (g), (h) to (j), (k) to (m) and (n) to (p) show the evolution of the Fermi surface topology due to electron-electron interaction of strengths $U=0.3$ eV, $U=0.5$ eV, $U=0.7$ eV and $U=3.0$ eV respectively for suitable chemical potentials above, close and below the corresponding $n_c$ in the same colour scheme for the six (renormalized) bands.} 
\label{fermi_surface}
\end{center}
\end{figure*}

%=============================================
Electrons in q2DEL predominantly occupy the three $t_{2g}$ orbitals of Ti ions at the terminating TiO$_2$ layer. These orbitals have parabolic dispersion near the $\Gamma$ point with the $d_{xy}$ orbital lying lower in energy than the $d_{xz}$, $d_{yz}$ orbitals by $0.4$ eV due to tetragonal distortion and quantum confinement at the interface~\cite{Held_PRB2013,Khalsa_PRB2013}. A small Rashba-like splitting is observed near the $\Gamma$ point while an atomic SOC splits the degenerate states near the crossing points of the orbitals resulting in an inter-orbital level mixing, as described in Fig.~\ref{fermi_surface}(a). Here, we study the normal state transport properties in the presence of electron correlation and ignore, for simplicity, the competing ferromagnetic and superconducting orders. We also ignore the real-space inhomogeneity at the interface because the effect of non-magnetic disorder is rather uninteresting in the context of the current study.

The non-interacting Hamiltonian, describing the interface electrons in the normal state, written in the basis of the three $t_{2g}$ orbitals, is given by~\cite{Held_PRB2013}
\begin{equation}
{\cal H}={\cal H}_0+{\cal H}_{ASO}+{\cal H}_{RSO}
\label{H_non_int}
\end{equation}
where ${\cal H}_0=\sum_{k,\alpha,\sigma}(\epsilon_{k\alpha}-\mu) c_{k\alpha\sigma}^\dagger c_{k\alpha\sigma}$ describes the band dispersion of the electrons in the three $t_{2g}$ orbitals with $\epsilon_{ka}=-2t_1(\cos{k_x}+\cos{k_y})-t_2-4t_3\cos{k_x}\cos{k_y}$, $\epsilon_{kb}=-t_1(1+2\cos{k_y})-2t_2\cos{k_x}-2t_3\cos{k_y}$, $\epsilon_{kc}=-t_1(1+2\cos{k_x})-2t_2\cos{k_y}-2t_3\cos{k_x}$, $\mu$ is the chemical potential and $t_1=0.277$ eV, $t_2=0.031$ eV, $t_3=0.076$ eV are the tight-binding parameters.

\begin{figure}[t]
\begin{center}
\epsfig{file=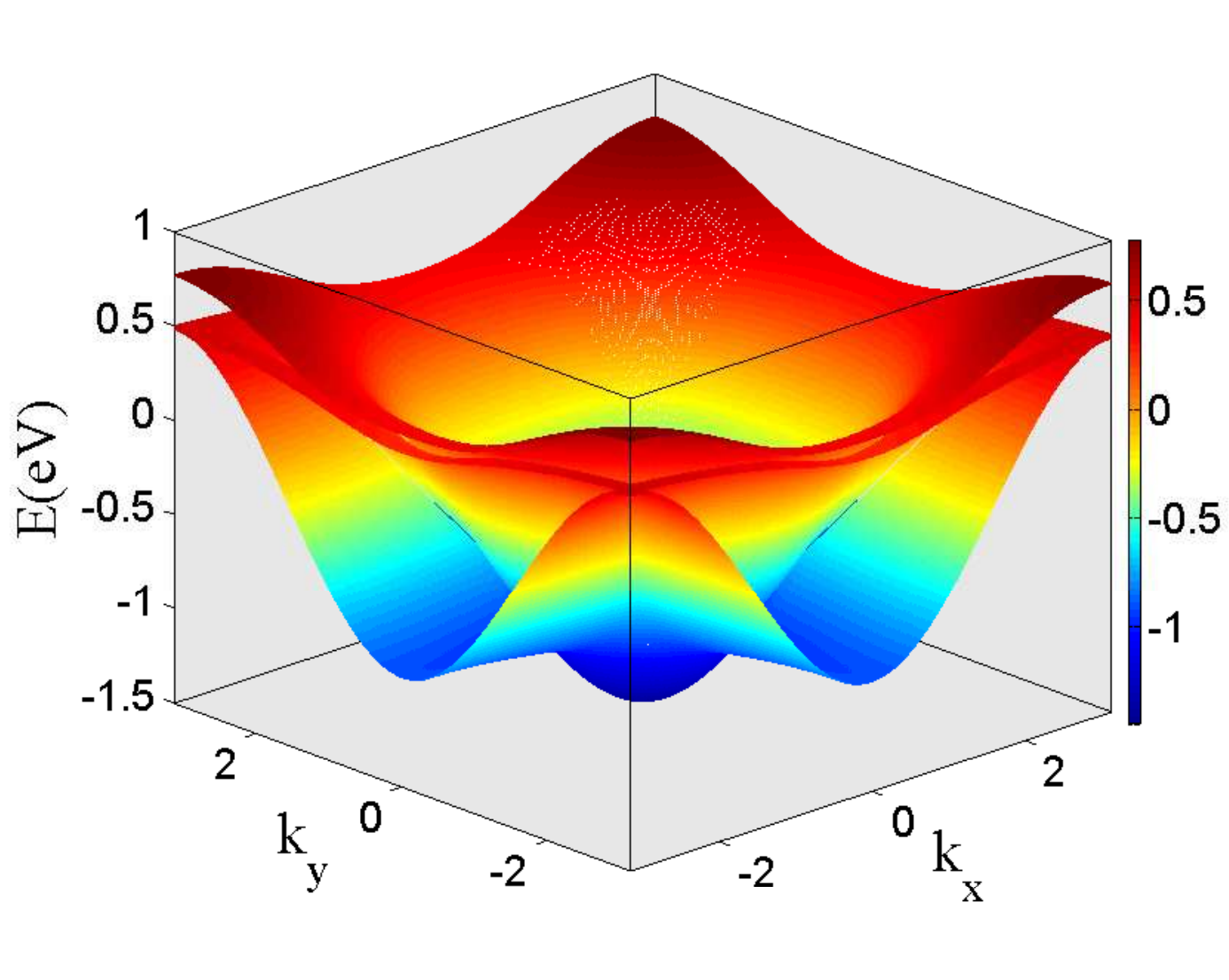,trim=0.0in 0.05in 0.0in 0.05in,clip=true, width=65mm}\vspace{0em}
\caption{(Color online) 3D band dispersion of the q2DEL at the interface obtained by diagonalizing the non-interacting part of the Hamiltonian.}
\label{3d_dispersion}
\end{center}
\end{figure}

The second term of the Hamiltonian is due to the atomic SOC ${\cal H}_{ASO}=\Delta_{so}\vec{l}\cdot\vec{s}$ which appears because of the crystal field splitting of the atomic orbitals. In terms of $t_{2g}$ orbital basis $(c_{ka\uparrow},c_{kb\uparrow},c_{kc\uparrow},c_{ka\downarrow},c_{kb\downarrow},c_{kc\downarrow})$ where ($a=d_{xy}$, $b=d_{yz}$, $c=d_{zx}$), ${\cal H}_{ASO}$ can be written as 
\begin{align}
&{\cal H}_{ASO}=\frac{\Delta_{so}}{2}\sum_{k}\begin{pmatrix} \begin{array}{cccccc} c_{ka\uparrow}^\dagger & c_{kb\uparrow}^\dagger & c_{kc\uparrow}^\dagger & c_{ka\downarrow}^\dagger & c_{kb\downarrow}^\dagger & c_{kc\downarrow}^\dagger\end{array} \end{pmatrix}\nonumber\\
&\times\begin{pmatrix} \begin{array}{cccccc} 0 & 0 & 0 & 0 & 1 & -i\\0 & 0 & i & -1 & 0 & 0\\0 & -i & 0 & i & 0 & 0\\0 & -1 & -i & 0 & 0 & 0\\1 & 0 & 0 & 0 & 0 & -i\\i & 0 & 0 & 0 & i & 0 \end{array} \end{pmatrix} \begin{pmatrix} \begin{array}{c} c_{ka\uparrow} \\ c_{kb\uparrow} \\ c_{kc\uparrow} \\ c_{ka\downarrow} \\ c_{kb\downarrow} \\ c_{kc\downarrow}\end{array} \end{pmatrix}
\label{H_aso}
\end{align}
where the strength of the atomic SOC is $\Delta_{so}=19.3$ meV. The third term of the Hamiltonian is due to the Rashba SOC, appearing because of the broken inversion symmetry at the interface, and is expressed as
\begin{align}
&{\cal H}_{RSO}=\gamma\sum_{k,\sigma}\begin{pmatrix} \begin{array}{ccc} c_{ka\sigma}^\dagger & c_{kb\sigma}^\dagger & c_{kc\sigma}^\dagger\end{array} \end{pmatrix}\nonumber\\
&\times\begin{pmatrix} \begin{array}{ccc} 0 & -2i\sin{k_x} & -2i\sin{k_y}\\2i\sin{k_x} & 0 & 0\\2i\sin{k_y} & 0 & 0\end{array} \end{pmatrix} \begin{pmatrix} \begin{array}{c} c_{ka\sigma} \\ c_{kb\sigma} \\ c_{kc\sigma}\end{array} \end{pmatrix}
\label{H_rso}
\end{align}
where $\gamma=20$ meV is the strength of the Rashba SOC. The 3D dispersion of all six bands of the q2DEL at the interface in the presence of the Rashba and atomic SOCs are shown in Fig.~\ref{3d_dispersion}. 
%=======================================================================================================================================================================
%=========================================================================================================================================================================
To treat the electron correlation in the q2DEL, we use a multi-orbital DMFT method in which the information of the non-interacting orbital is fed through the density of states of each individual orbital. But in the non-interacting three-orbital Hamiltonian in Eq.~(\ref{H_non_int}), the spin and orbital degrees of freedom are entangled due to the presence of the SOCs resulting in an effective six orbital system where electron spin is no longer a good quantum number. Therefore, we describe the electron-electron interaction using a six-orbital Hubbard Hamiltonian given by

%--------------------------------------------------------------------------------------------------------------------------------------------------------------------------------
%--------------------------------------------------------------------------------------------------------------------------------------------------------------------------------

\begin{align}
&{\cal H}_{U}=\sum_{\alpha,\alpha'}U_{\alpha\alpha^{\prime}} n_{\alpha}n_{\alpha^{\prime}}+\sum_{\alpha,\beta}U_{\alpha\beta} n_{\alpha}n_{\beta}
\label{H_U}
\end{align}
\noindent where $U_{\alpha\alpha^{\prime}}$ is the repulsive interaction energies between electrons in the pairs of nearly-degenerate orbitals with orbital indices $\{ \alpha=1,3,5$, $\alpha^{\prime}=\alpha +1 \}$ while $U_{\alpha\beta}$ is the interaction energies between electrons in the widely separated orbitals \textit{i.e.} $\{ \alpha=1,2$, $\beta=3,4 \}$ and $\{ \alpha=3,4$, $\beta=5,6 \}$. Since the interaction between electrons in the nearly-degenerate orbitals is stronger than that between electrons in the widely separated orbitals, we take $U_{\alpha\beta}=\frac{U_{\alpha\alpha^{\prime}}}{5}$. For notational convenience, we shall therefore use $U=U_{\alpha\alpha^{\prime}}$ in the rest of the paper. The total occupation number is computed using $n=\sum_{\alpha=1}^{6} \langle c_{\alpha}^{\dagger}c_{\alpha} \rangle$. We solve the effective Hamiltonian ${\cal H}_{eff}={\cal H}+{\cal H}_{U}$ for different values of $U$ ranging from $0$ to $0.7$ eV and for different carrier concentration across the Lifshitz transition. There are, however, few reports of the values of these  correlations, $U$ and $U_{\alpha\beta}$. There is so far no indication of strong correlation in this system from experimental observations~\cite{ncomm_nagaosa}. Whereas a DFT calculation~\cite{imada} puts the interface electrons in the strong coupling limit, $U \simeq 3.6$ eV and $U_{\alpha\beta}=U/1.3$, others report a much lower value $U < 1$ eV~\cite{ncomm_nagaosa}. In view of such disparities, it is useful to check the results at both intermediate and strong coupling limits which we report in the foregoing. We show that the qualitative features remain quite similar in both the limits, while the critical electron density for Lifshitz transition changes with changing correlations. Since the electron-density is very low here, the effect of correlation is not expected to be dramatic as in a prototypical correlated electronic system. We work in the weak to intermediate interaction regime, with maximum $U$ value upto 0.7 eV, where the typical band-width here is about 2 eV. As mentioned above, we also compare the results in the prescribed strong coupling limit of the DFT calculation~\cite{imada}. 

%=============================================
\section{Results}
\label{results}
%=============================================
\subsection{Correlated Fermi Surface}
\label{Correlated Fermi Surface}
%=============================================

The Lifshitz transition refers to the change in topology of the Fermi surface (e.g., formation or disappearance of a pocket or an arc) by tuning of some parameter~\cite{Imada_JPSJ} which, in the present case, is the chemical potential or the carrier density. In the experimental set up, usualy the gate voltage is varied to tune carrier density. From our data, it is quite easy to invert the carrier densities to the corresponding gate voltages~\cite{PhysRevB.92.174531}. The non-interacting band structure and the Fermi surfaces at three different carrier densities are shown in Fig.~\ref{fermi_surface}(a)-(d). At low carrier densities the two lower orbitals are occupied resulting in a pair of elliptical Fermi surfaces. As the chemical potential is tuned up, new electron-like pockets appear at a critical value $\mu_c=-1.067$ eV (corresponding carrier density $n_c=0.1809$ el./u.c.), where the 3$^{rd}$ and 4$^{th}$ orbitals start getting occupied. The orbitals 1$^{st}$ and 2$^{nd}$, 3$^{rd}$ and 4$^{th}$, 5$^{th}$ and 6$^{th}$ are pairwise nearly degenerate at the $\Gamma$ point and there exists another critical density at which the 5$^{th}$ and 6$^{th}$ orbitals will appear at the Fermi level and new pockets will form. The difference between these two critical densities is quite small in the present case to distinguish the first Lifshitz transition from the second. However, if such an interface is availble where the atomic SOC is large this difference would be considerable and clearly delineable~\cite{note1}. 

To get an impression of the effect of electron-electron interaction on the Fermi surface and hence on the Lifshitz transition, we calculate the correlated Fermi surfaces and compare with the non-interacting case. In the presence of electron correlations, the renormalized dispersion $E_{k\alpha}$ for orbital $\alpha$ is given by
\begin{equation*}
E_{k\alpha}=E_{k\alpha}^{0}-\mu+Re \Sigma_{\alpha} (\mathbf{k},E_{k\alpha})
\end{equation*}
where $E_{k\alpha}^{0}$ is the non-interacting dispersion, obtained by diagonalizing the Hamiltonian~(\ref{H_non_int}) at each momentum point $\mathbf{k}$, and $\Sigma_{\alpha}(\mathbf{k},E_{k\alpha})$ is the self-energy. Fig.~\ref{fermi_surface}(e)-(p) show the correlated Fermi surfaces for various interaction strengths and carrier densities across the Lifshitz transition. With increasing $U$, the spectral weight near the $\Gamma$ point is transferred towards lower energy and the nearly degenerate pairs of orbitals tend to split up. As shown in Fig.~\ref{fermi_surface}(l), at larger correlation strength, $U=0.7$ eV, the 5$^{th}$ and 6$^{th}$ orbitals also appear as new electron-like pockets in the Fermi surface, indicating the second Lifshitz transition driven by electron correlations. Similar progression of Fermi surface topology happens when the chemical potential is suitably tuned in the strong correlation limit ($U=3$ eV, Fig.~\ref{fermi_surface}(n)-(p)). This essentially means that the critical densities for the transitions are reduced when interaction increases. If we look away from the critical point, we see that while there is no noticeable change due to increasing interaction in Fig.~\ref{fermi_surface}(g),(j),(m), (p) for the lower two orbitals, Fig.~\ref{fermi_surface}(e),(h),(k), (n) for the upper four orbitals show prominent changes. 

%=============================================
\subsection{Transport properties}
\label{transport}
%=============================================
To study the influence of the Lifshitz transition on the transport, various transport quantities are calculated. The transport coefficients that govern the electrical and thermal responses of the system are given in terms of current-current correlation functions which reduce to averages over the spectral density $\rho(\epsilon,\omega)$ in DMFT.
The expressions for optical conductivity ($\sigma(\omega)$), Seebeck coefficient(S), and thermal conductivity($K$) are ~\cite{Kotliar_PRL1998} 

\begin{figure}[t]
\begin{center}
\epsfig{file=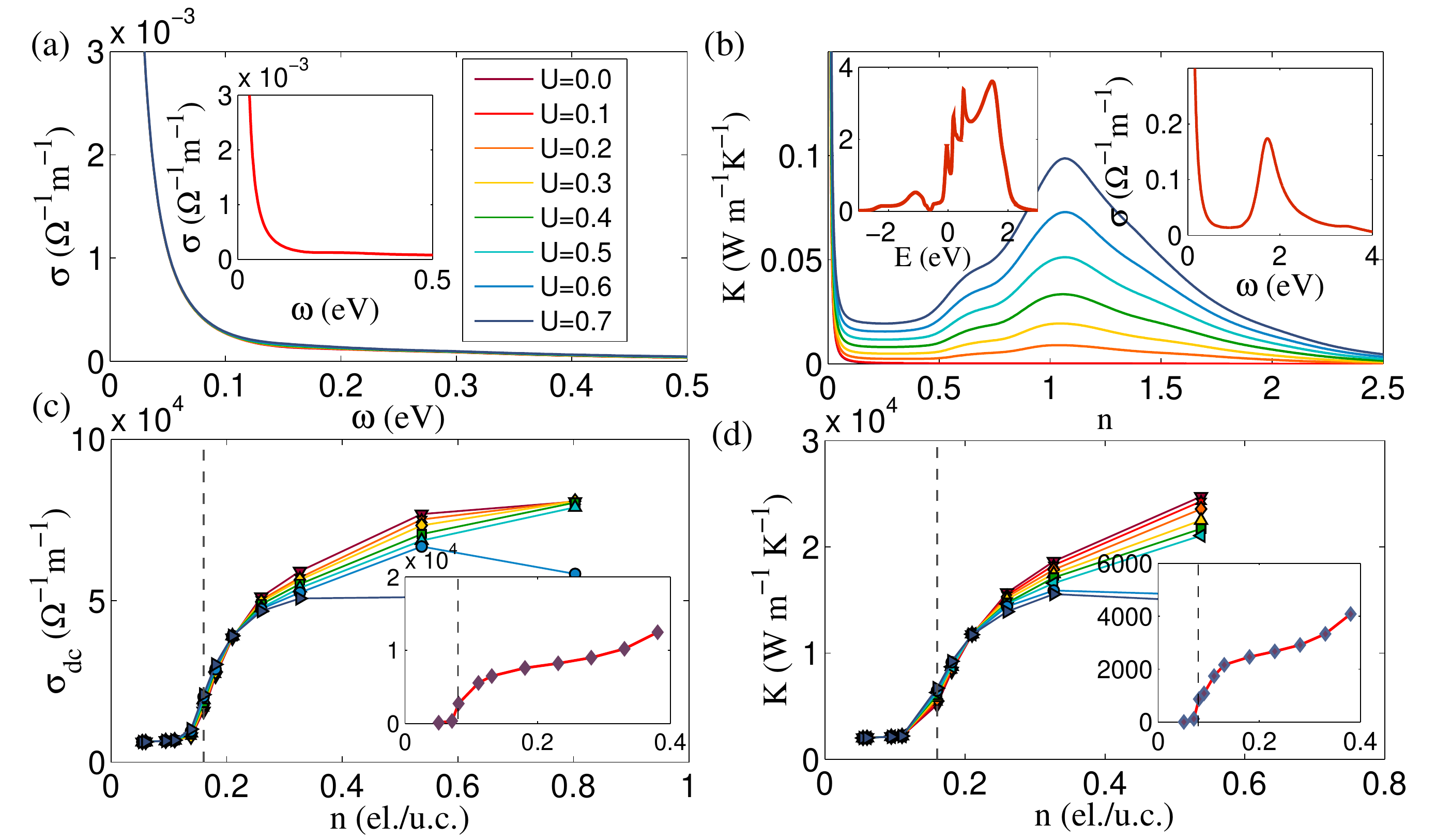,trim=0.1in 0.05in 0.0in 0.05in,clip=true, width=90mm}\vspace{0em}
\caption{(Color online) The optical conductivity as a function of energy at 60 K for several values of interaction strength $U$ with carrier concentration (a) $n$ below Lifshitz transition (inset, for $U=3$ eV) and (b) above Lifshitz transition (right inset, $U=3$ eV; the corresponding DOS (in arb. units) in the left inset). The (c) dc conductivity and (d) thermal conductivity as a function of the carrier concentration with $U$ (in eV) same as in (a) and (b), showing the rapid rise beyond critical carrier concentration, marked by dashed line, $n_c$ ($U=0$).}
\label{optical}
\end{center}
\end{figure}

\begin{center}
$\sigma=\frac{e^{2}}{T}A_{0}$, $S=\frac{-k_{B}}{e}\frac{A_{1}}{A_{0}}$,  $K=k_{B}^{2}(A_{2}-\frac{A^{2}_{1}}{A_{0}})$
\end{center}
where 
\begin{equation*}
A_{n} = \frac{\pi}{\hbar V}\sum_{k,\sigma}\int d\omega \rho_{\sigma}(k,\omega)^{2}(\frac{\partial{\epsilon_{k}}}{\partial{k_{x}}})^{2} (-T \frac{\partial{f(\omega)}}{\partial{\omega}}) (\beta \omega)^{n}
\end{equation*}
Here V is the volume and $f(\omega)$ is Fermi-Dirac distribution function.
Figs.~\ref{optical}(a) and (b) show the optical conductivity $\sigma(\omega)$ for different values of the electron-electron interaction strength $U$ and the carrier concentration $n$, respectively, below and above the critical concentration $n_c$. At $n<n_c$, $\sigma(\omega)$ follows a hyperbolic dependence with energy $\omega$ showing insignificant effects of the interaction. On the other hand, at $n>n_c$, $\sigma(\omega)$ reveals, at lower $U$, shoulder-like feature which transforms into a peak near $\omega_c\simeq1.0$ eV with increasing $U$.
Since the effective non-interacting band-width for the system is about 2 eV, a Hubbard correlation strength of 3 eV transfers the spectral weight to higher energies and an upper Hubbard band starts forming, as shown in the left inset of Fig.~\ref{optical}(b). The signature of the same appears in the optical conductivity where a strong sub-peak at energies $\sim 1.8-2.2$ eV forms (right inset of Fig.~\ref{optical}(b)), suppressing the secondary small peak structures observed for the lower $U$ values. The Drude peak gets suppressed significantly as well, consistent with the increment in resistivity (decrease in DC conductivity) at $U=3.0$ eV.
%----------------------------------------------------------------------------------------------------------------------------------------------
%----------------------------------------------------------------------------------------------------------------------------------------------
%----------------------------------------------------------------------------------------------------------------------------------------------
\begin{figure}[t]
\begin{center}
\epsfig{file=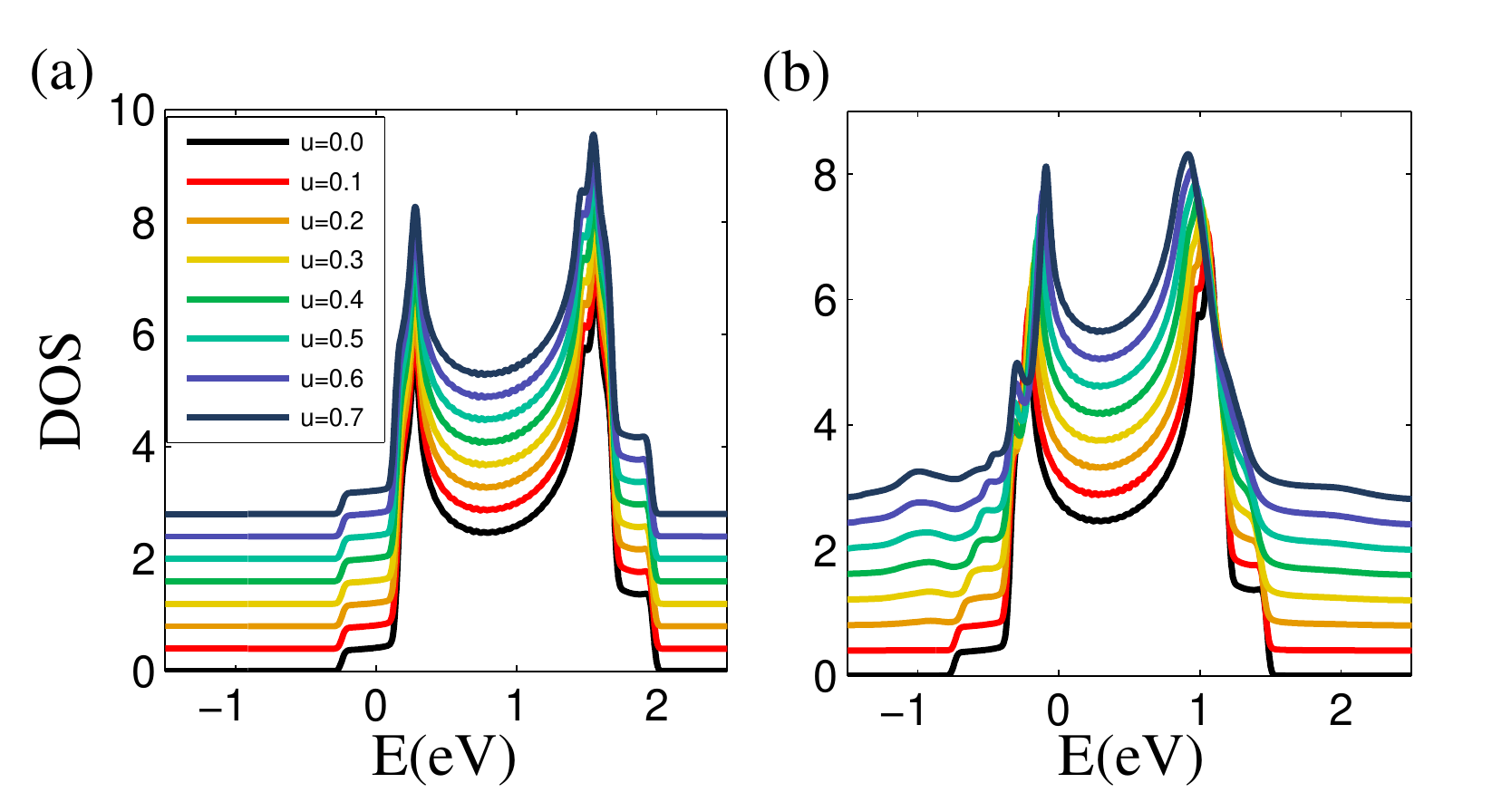,trim=0.10in 0.15in 0.in 0.0in,clip=true, width=90mm}\vspace{0em}
\caption{(Color online) Total density of states as a function of energy at 60 K for several values of $U$ with carrier concentration (a) $n=0.0945$ el./u.c. (below the Lifshitz transition)  and (b) $n=1.4030$ el./u.c. (above Lifshitz transition). The curves from $U=0.1$ eV to $U=0.7$ eV are progressively offset vertically by 0.4 for clarity of viewing, i.e., $U=0.1$ eV is shifted up by 0.4 and $U=0.7$ eV by 0.24.}
\label{dos}
\end{center}
\end{figure}
%----------------------------------------------------------------------------------------------------------------------------------------------
The increasing peak height of $\sigma(\omega_c)$ with increasing $U$ indicates transfer of spectral weight toward higher energies. A closer look into $\sigma(\omega)$ reveals a two satellite structure of optical conductivity with satellites around 0.5 eV and 1.0 eV respectively. This is again consistent with the features of correlated spectral weight transfer in the net density of states (DOS) for the system as shown in Fig.~\ref{dos}. The edge singularities of the DOS arise from the nearly non-dispersive segments of the band structure. When correlation is cranked up, finite spectral weight transfers from the edge singularities lead to satellite features at $\omega$ $\sim 0.5$ eV and $1.0$ eV respectively. As described in Figs.~\ref{optical} (c) and (d), the dc conductivity ($\sigma_{dc}= \sigma(\omega=0)$) and the thermal conductivity $K$ rise rapidly to higher values with increasing carrier concentration $n$. This sharp change arises because a large number of accessible states appear at the Fermi level from the newly occupied orbitals and due to the nearly flat nature of the two lower orbitals above $n_c$. Evidently, at higher $n$, both the conductivities reduce with increasing $U$ as a consequence of the piling up of the spectral weight at $\omega_c$ as discussed before in Fig.~\ref{optical}(b). The strong-coupling limit (inset, Fig.~\ref{optical}) is quite featureless, a considerable spectral weight is now transferred to the high energy region with a peak corresponding to the strong interaction.
%----------------------------------------------------------------------------------------------------------------------------------------------
\begin{figure}[t]
\begin{center}
\epsfig{file=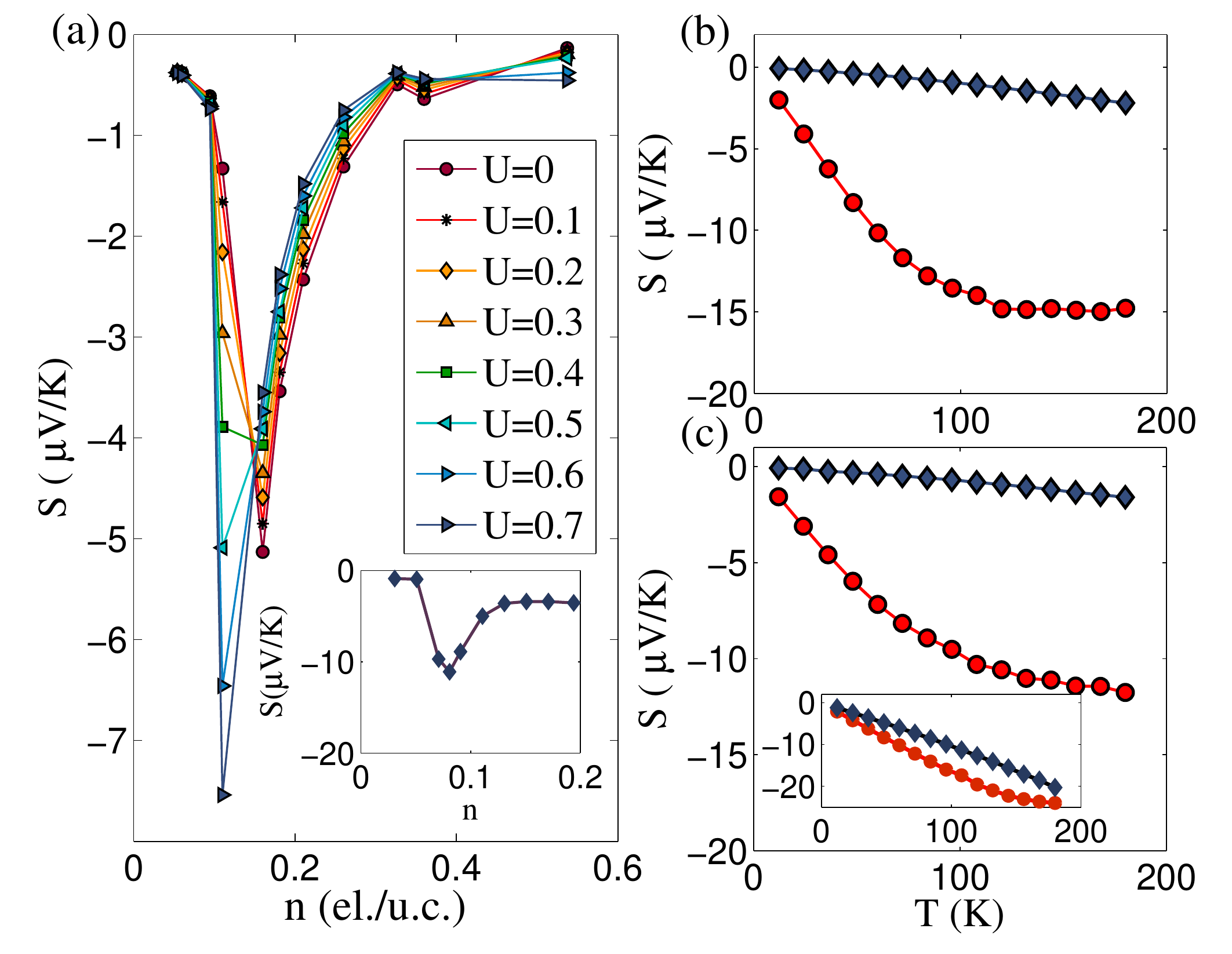,trim=0.10in 0.15in 0.in 0.0in,clip=true, width=90mm}\vspace{0em}
\caption{(Color online)  (a) The Seebeck coefficient $S$ as a function of the electron concentration $n$ at a temperature $T=60$ K for various strengths $U$ (in eV unit) of the electron-electron interaction showing the cusp at the critical carrier concentration $n_c$ (inset, $U=3$ eV). Variation of $S$ with respect to temperature $T$ below critical carrier concentration (red) and above critical carrier concentration (blue) for (b) $U=0.3$ eV and (c) $U=0.7$ eV, respectively. (inset, $U=3$ eV)}
\label{thermo1}
\end{center}
\end{figure}
%----------------------------------------------------------------------------------------------------------------------------------------------
The thermopower or the Seebeck coefficient $S$, as depicted in Fig.~\ref{thermo1}(a), reveals a cusp at the critical carrier concentration $n_c$ for the Lifshitz transition. With the enhancement in $U$, the cusp shifts largely towards lower $n$ values, referred to above from the plots of the correlated Fermi surfaces; the depth of the cusp also increases. Fig.~\ref{thermo1}(b), (c) show the variation of $S$ with respect to temperature $T$ at carrier densities above and below the critical carrier density for $U=0.3$ eV and $U=0.7$ eV, respectively. The same variation is also observed for strong coupling case (inset, Fig.~\ref{thermo1}). The Seebeck coefficient varies almost linearly with temperature at $n>n_{c}$ However, at $n < n_{c}$, $S$ decreases with temperature below $100$ K and then becomes nearly flat. With increasing $U$, $S$ can increase or decrease depending on the exact value of $n$ as noticed in Fig.~\ref{thermo1}(a). Very similar temperature dependence of the Seebeck coefficient has been reported in the experiment~\cite{Triscone_PRB2010}.

%=============================================
\subsection{Theoretical ARPES}
\label{arpes}
%=============================================
To get more insight into the spectral weight transfer occurring in the presence of finite electron correlations, we study the ARPES intensity spectrum. The spectral intensity can be expressed as~\cite{Valla_PRL2000}, where $\Sigma(\omega)$ is the momentum-independent self-energy from DMFT
\begin{equation*}
I(\bold{k}, \omega) \propto \frac{Im\Sigma(\omega) f(\omega)}{[\omega-\epsilon_{k}-Re\Sigma(\omega)]^{2} + [Im\Sigma(\omega)]^{2}} 
\end{equation*}
where $f(\omega)$ is the Fermi-Dirac distribution function. 
%----------------------------------------------------------------------------------------------------------------------------------------------
\begin{figure*}[t]
\begin{center}
\epsfig{file=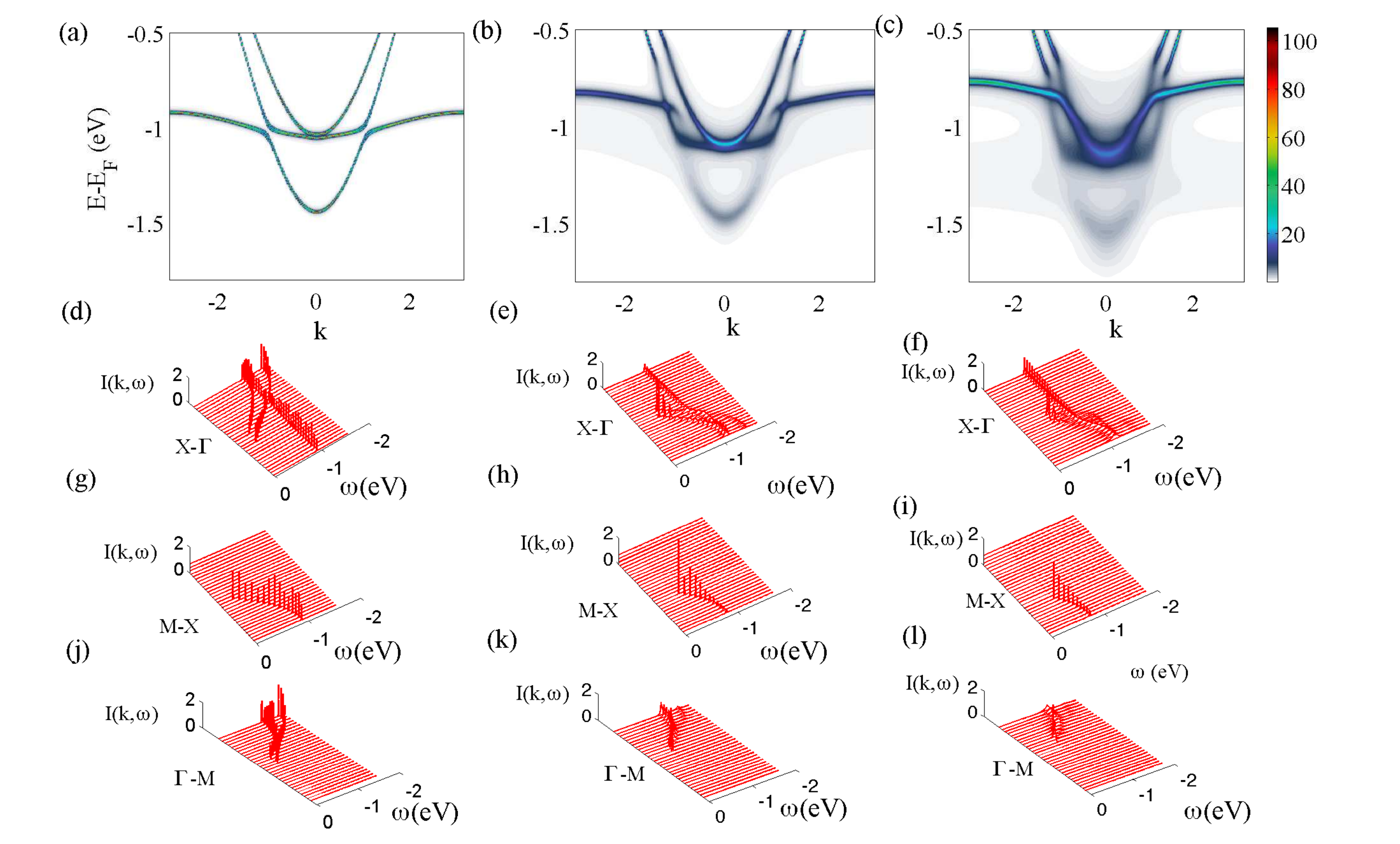,trim=0.4in 0.05in 0.0in 0.1in,clip=true, width=180mm}\vspace{0em}
\caption{(Color online) The ARPES intensity spectra along $\bar{X}$-$\Gamma$-X direction (in units of $\pi$) for (a) $U=0$, (b) $U=0.3$ eV and (c) $U=0.5$ eV, respectively at a carrier concentration $n=1.4030$ el./u.c. and at a fixed temperature $T=60$ K. (d)-(l) show the momentum-resolved ARPES intensity along $\Gamma$-X, X-M and M-$\Gamma$ directions in the Brillouin zone (rows) and for $U=0$, $U=0.3$ eV and (c) $U=0.5$ eV (columns) at a fixed temperature $T=60$ K.}
\label{arpes_f}
\end{center}
\end{figure*}
%----------------------------------------------------------------------------------------------------------------------------------------------
Figs.~\ref{arpes_f}(a), (b), (c), show and compare the ARPES spectra at finite interaction strength $U$ with that of the non-interacting case. The ARPES has the highest intensity and spectral weight around the high symmetry $\Gamma$ point at the non-interacting level. In case of orbitals with predominantly $d_{xz}$ and $d_{yz}$ characters, the intense segments in the ARPES are due to the nearly non-dispersive sections of the orbitals with heavy mass. At the non-interacting level, this heavy mass is a direct consequence of finite atomic spin-orbit coupling. With correlation the weight transfers over finite energy window. For the $d_{xy}$ orbital the spectral weight gets transferred over an energy range of $\sim 0.15$ eV and $\sim 0.3$ eV for $U=0.3$ eV and $0.5$ eV respectively around the $\Gamma$ point. Around the same point for $d_{xz}$ and $d_{yz}$ orbitals the spectral weights get transferred over an energy scale of $\sim 0.2$ eV and $\sim 0.4$ eV for $U=0.3$ eV and 0.5 eV respectively. As the spectral weight transfers with finite correlation, the ARPES intensity profile around the $\Gamma$ point becomes significantly weaker; for some orbitals intensity drops by a factor of ten with $U=\frac{W}{4}$ (where W is the effective bandwidth of the dispersive bands). The spectral weight at the region near the $\Gamma$ point is transferred towards lower energies while that at the regions away from the $\Gamma$ point is transferred towards higher energies, thereby increasing the bandwidth of each orbital. Higher the $U$, larger is the spectral weight transfer. The plots in Figs.~\ref{arpes_f}(d)-(l) describe the ARPES intensity along high-symmetry directions in the Brillouin zone. The inter-orbital finite interaction term at the local site within DMFT, pulls the bands of predominantly $d_{xy}$ and $d_{xz,yz}$ characters towards each other (pulling the $d_{xy}$ bands toward smaller energies and driving the $d_{xz,yz}$ bands toward higher energies) allowing them to overlap and admix substantially. 

%=============================================
\subsection{Influence of magnetic field}
\label{field}
\begin{figure}[t]
\begin{center}
\epsfig{file=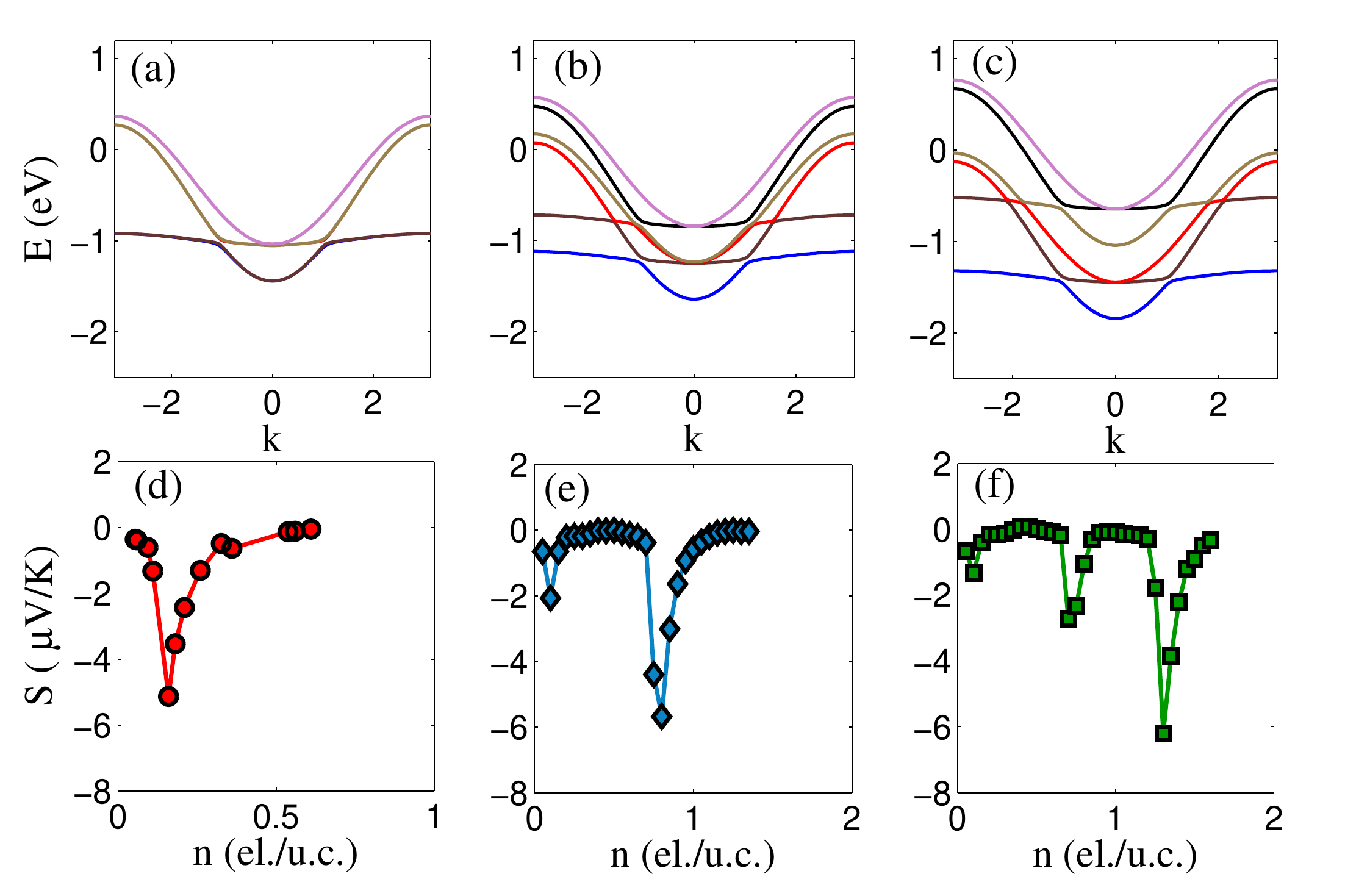,trim=0.10in 0.15in 0.0in 0.0in,clip=true, width=90mm}
\caption{(Color online) The non-interacting band structure $\bar{X}$-$\Gamma$-X direction (in units of $\pi$) with in-plane magnetic field of strength (a) $h_{x}=0$, (b) $h_{x}=0.2$ eV and (c) $h_{x}=0.4$ eV. (d)-(f) are the corresponding plots of the Seebeck coefficients as a function of the carrier concentration at temperature $T=60$ K.}
\label{with_field}
\end{center}
\end{figure}

\begin{figure}[t]
\begin{center}
\epsfig{file=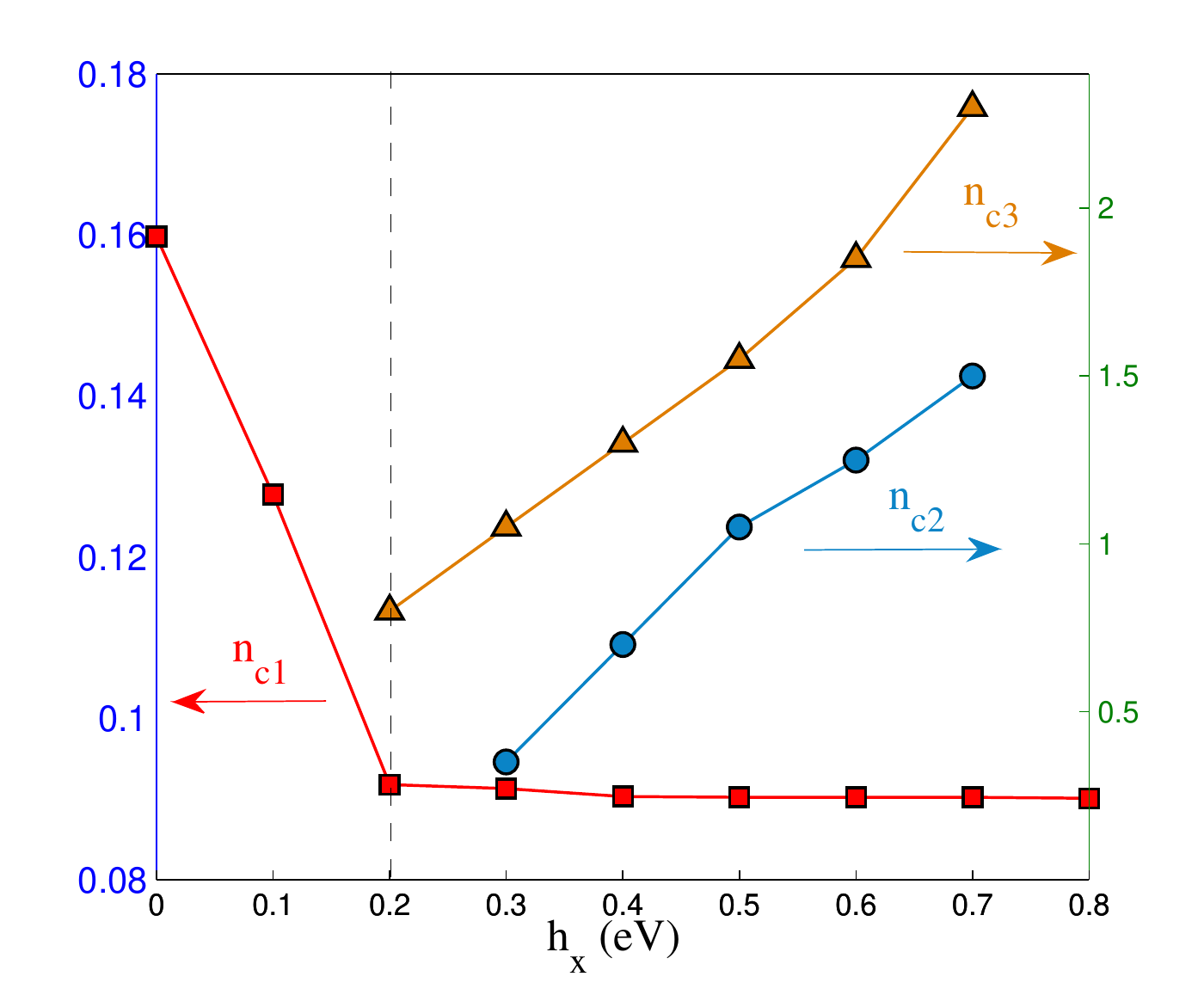,trim=0.15in 0.08in 0.0in 0.0in,clip=true, width=80mm}
\caption{(Color online) The critical density for the first ($n_{c1}$), second ($n_{c2}$) and third ($n_{c3}$) Lifshitz transitions as a function of the strength $h_{x}$ of the in-plane magnetic field. Left y-axis label is for $n_{c1}$ and right y-axis label is for $n_{c2}$ and $n_{c3}$. The numbers in both the y-axes are in the unit of el./u.c.}
\label{hx_nc}
\end{center}
\end{figure}

%=============================================
Having explored the basic features of the Lifshitz transition and the role of the electron correlations in it, we study the effects with an external magnetic field. Figs.~\ref{with_field}(a)-(c), show the band structure with an exchange field applied along the x-direction in the absence of any electron correlations. With a finite field, the degeneracy is lifted at all momenta, except at the $\Gamma$ point, creating a gap. However, the exchange split orbitals undergo band inversion at all the band crossing points due to the presence of the SOCs.      
%------------------------------------------------------------------------------------------
%------------------------------------------------------------------------------------------
Figs.~\ref{with_field}(a)-(c) describe the corresponding Seebeck coefficients as a function of the carrier concentration. In addition to the cusp in the zero-field case, other cusps appear at larger fields indicating multiple Lifshitz transitions. The appearance of the multiple Lifshitz transition can be reconciled from the band structures directly. With the tuning of the Fermi level up from the two lower orbitals, new electron-like pockets appear at all the carrier concentrations for which the Fermi level encounters new orbital contribution at the $\Gamma$ point. The degeneracy at the $\Gamma$ point is maintained even at higher fields. For multiple transitions, the Seebeck coefficient at the higher (at larger $n$) cusps are larger in magnitude. For a quantitative description, the variation of the critical carrier densities for these Lifshitz transitions, given by the locations of the cusps in the Seebeck coefficient, with respect to the field strength $h_x$ is shown in Fig.~\ref{hx_nc}.
%------------------------------------------------------------------------------------------
%------------------------------------------------------------------------------------------
There are two clear regimes of field strength: the low field regime ($h_x<0.2$ eV) has only one Lifshitz transition whereas multiple transitions appear in higher fields ($h_x \geqslant 0.2$ eV). The critical density $n_{c1}$ for the first transition decreases with increasing $h_x$, as observed in the experiment~\cite{Joshua_PNAS2013}, and then becomes nearly constant with further increase in $h_x$ in the second field regime. On the other hand, the second and third critical concentrations $n_{c2}$ and $n_{c3}$, respectively, increase with increasing $h_x$. With the limited resolution, the data suggest that $n_{c2}$ and $n_{c3}$ originally coincide with $n_{c1}$ at lower fields and get separated with increasing $h_x$ due to the removed degeneracy at the $\Gamma$ point. The third Lifshitz transition occurs at lower fields than the second transition, as evident from Fig.~\ref{with_field}. The decrease/increase in the critical density/densities in the low/high field regime is a manifestation of the fact that the Zeeman field splits and pushes the bands away from each other.  We also study the case with field applied perpendicular to the interface, and find similar qualitative features of the multiple Lifshitz transitions. 

%=============================================
\section{Discussion and Summary}
\label{summary}
%=============================================
An experimentally testable scenario involving the Lifshitz transition in the LAO/STO interface has been worked out. The transport properties of the q2DEL provide important signatures of the Lifshitz transition. The cusp in the Seebeck coefficient is a robust response of the change in the Fermi surface topology and can be used to examine Lifshitz transition in metallic systems. The electron-electron interaction may result in a Pomeranchuk instability in similar multi-orbital materials. External magnetic field can lead to multiple Lifshitz transitions which shows up as multiple cusps in the Seebeck coefficient. However, detecting the multiple transitions in the experiment will require scanning over large exchange energies. Also, the carrier density in the LAO/STO hetero-interface is small compared to other metallic interfaces and, therefore, difficult to tune to large values using the external gate-voltage. Nevertheless, the phenomena of multiple Lifshitz transitions is possible in other similar multi-orbital metallic systems and requires intensive effort to realize in the experiment.

To summarize, we have studied the Lifshitz transition in the LAO/STO interface and its influence on the transport properties using multi-orbital DMFT, formulated within realistic bands at the interface. Though it is still not established if the q2DEL belongs in the intermediate or strong correlation class, the effects of correlation are visible in the dynamical spectral features of the electron density even in the intermediate regime. In particular, the location of the Lifshitz transition and therefore the transport bear signatures of it. The Lifshitz transition occurs when new carriers from the $d_{yz}$ and $d_{zx}$ orbitals populate the Fermi level and new electron-like pockets appear at the Fermi surface. The change in the Fermi surface topology appears as a sharp rise in the dc and thermal conductivities and as a cusp in the Seebeck coefficient. The repulsive electron-electron interaction transfers some of the spectral weights towards lower energies and effectively reduces the critical carrier concentration for the transition. The pronounced cusp features in the Seebeck coefficient can be a very handy tool for the identification of the Lifshitz transition, otherwise located by tracking the carrier concentration, not always very reliable in the oxide interfaces. In the presence of external magnetic field, the critical carrier concentration further reduces with increasing field strength and multiple transitions appear at sufficiently large fields.

%=============================================
\section*{Acknowledgements}
%=============================================
The authors thank Siddhartha Lal for discussions and acknowledge the use of the computing facility from DST-Fund for Improvement of S~$\&$~T infrastructure (phase-II) Project installed in the Department of Physics, IIT Kharagpur, India. SN acknowledges MHRD, India for support. NM was supported by the DFG through TRR 80. SA acknowledges UGC, India for a research fellowship.
%=============================================

\bibliographystyle{h-physrev}
%\bibliography{interface}

\begin{thebibliography}{10}

\bibitem{Ohtomo2004}
A.~Ohtomo and H.~Y. Hwang,
\newblock Nature {\bf 427}, 423 (2004).

\bibitem{Thiel_Science2006}
S.~Thiel and G.~Hammerl and A.~Schmehl and C.~W. Schneider and J.~Mannhart,
\newblock Science {\bf 313}, 1942 (2006).

\bibitem{Huijben_NMat2006}
M.~Huijben and G.~Rijnders and D.~H.~A. Blank and S.~Bals and S.~V. Aert and
  J.~Verbeeck and G.~V. Tendeloo and A.~Brinkman and H.~Hilgenkamp,
\newblock Nat. Mater. {\bf 5}, 556 (2006).

\bibitem{Reyren31082007}
N.~Reyren, S.~Thiel, A.~D. Caviglia, L.~F. Kourkoutis and G.~Hammerl and
  C.~Richter and C.~W. Schneider and T.~Kopp and A.-S. Rüetschi and D.~Jaccard
  and M.~Gabay and D.~A. Muller and J.-M. Triscone and J.~Mannhart,
\newblock Science {\bf 317}, 1196 (2007).

\bibitem{GariglioJPCM2009}
S.~Gariglio and N.~Reyren and A.~D. Caviglia and J.-M. Triscone,
\newblock J. Phys.: Condens. Matter {\bf 21}, 164213 (2009).

\bibitem{RichterNature2013}
C.~Richter and H.~Boschker and W.~Dietsche and E.~Fillis-Tsirakis and R.~Jany
  and F.~Loder and L.~F. Kourkoutis and D.~A. Muller and J.~R. Kirtley and
  C.~W. Schneider and J.~Mannhart,
\newblock Nature {\bf 502}, 528 (2013).

\bibitem{Marel_PRB2014}
S.~N. Klimin and J.~Tempere and J.~T. Devreese and D.~van~der Marel,
\newblock Phys. Rev. B {\bf 89}, 184514 (2014).

\bibitem{Li_Nature2011}
L.~Li and C.~Richter and J.~Mannhart and R.~C. Ashoori,
\newblock Nat. Phys. {\bf 7}, 762 (2011).

\bibitem{Dikin_PRL2011}
D.~A. Dikin and M.~Mehta and C.~W. Bark and C.~M. Folkman and C.~B. Eom and
  V.~Chandrasekhar,
\newblock Phys. Rev. Lett. {\bf 107}, 056802 (2011).

\bibitem{Dagan_PRL2014}
A.~Ron and E.~Maniv and D.~Graf and J.-H. Park and Y.~Dagan,
\newblock Phys. Rev. Lett. {\bf 113}, 216801 (2014).

\bibitem{Tra_AdMa2013}
V.~T. Tra, J.-W. Chen, P.-C. Huang, B.-C. Huang, Y.~Cao, C.-H. Yeh, H.-J. Liu
  and E.~A. Eliseev and A.~N. Morozovska and J.-Y. Lin and Y.-C. Chen and M.-W.
  Chu and P.-W. Chiu and Y.-P. Chiu and L.-Q. Chen and C.-L. Wu and Y.-H. Chu,
\newblock Adv. Mater. {\bf 25}, 3357 (2013).

\bibitem{Fischer_NJP2013}
M.~H. Fischer and S.~Raghu and E.-A. Kim,
\newblock New J. Phys. {\bf 15}, 023022 (2013).

\bibitem{Fete_PRB2012}
A.~F\^ete and S.~Gariglio and A.~D. Caviglia and J.-M. Triscone and M.~Gabay,
\newblock Phys. Rev. B {\bf 86}, 201105 (2012).

\bibitem{Caviglia2008}
A.~D. Caviglia and S.~Gariglio and N.~Reyren and D.~Jaccard and T.~Schneider
  and M.~Gabay and S.~Thiel and G.~Hammerl and J.~Mannhart and J.-M. Triscone,
\newblock Nature {\bf 456}, 624 (2008).

\bibitem{CavigliaPRL2010}
A.~D. Caviglia and M.~Gabay and S.~Gariglio and N.~Reyren and C.~Cancellieri
  and J.-M. Triscone,
\newblock Phys. Rev. Lett. {\bf 104}, 126803 (2010).

\bibitem{Cen_NMat2008}
C.~Cen and S.~Thiel and G.~Hammerl and C.~W. Schneider and K.~E. Andersen and
  C.~S. Hellberg and J.~Mannhart and J.~Levy,
\newblock Nat. Mater. {\bf 7}, 298 (2008).

\bibitem{Balatsky_PRB2012}
J.~T. Haraldsen and P.~W\"olfle and A.~V. Balatsky,
\newblock Phys. Rev. B {\bf 85}, 134501 (2012).

\bibitem{Bucheli_PRB2014}
D.~Bucheli and M.~Grilli and F.~Peronaci and G.~Seibold and S.~Caprara,
\newblock Phys. Rev. B {\bf 89}, 195448 (2014).

\bibitem{Nakagawa2006}
N.~Nakagawa and H.~Y. Hwang and D.~A. Muller,
\newblock Nat. Mater. {\bf 5}, 204 (2006).

\bibitem{Dagotto1076}
E.~Dagotto,
\newblock Science {\bf 318}, 1076 (2007).

\bibitem{Rijnders_Nature2008}
G.~Rijnders and D.~H.~A. Blank,
\newblock Nat. Mater. {\bf 7}, 270 (2008).

\bibitem{Satpathy_PRL2008}
Z.~S. Popovi\ifmmode~\acute{c}\else \'{c}\fi{} and S.~Satpathy and R.~M.
  Martin,
\newblock Phys. Rev. Lett. {\bf 101}, 256801 (2008).

\bibitem{Ariando_PRX2013}
Z.~Q. Liu, C.~J. Li, W.~M. L\"u and X.~H. Huang and Z.~Huang and S.~W. Zeng and
  X.~P. Qiu and L.~S. Huang and A.~Annadi and J.~S. Chen and J.~M.~D. Coey and
  T.~Venkatesan and Ariando,
\newblock Phys. Rev. X {\bf 3}, 021010 (2013).

\bibitem{Li_PRB2011}
Y.~Li and S.~N. Phattalung and S.~Limpijumnong and J.~Kim and J.~Yu,
\newblock Phys. Rev. B {\bf 84}, 245307 (2011).

\bibitem{Zhong_PRB2010}
Z.~Zhong and P.~X. Xu and P.~J. Kelly,
\newblock Phys. Rev. B {\bf 82}, 165127 (2010).

\bibitem{Boschker_SciRep2015}
H.~Boschker and C.~Richter and E.~Fillis-Tsirakis and C.~W. Schneider and
  J.~Mannhart,
\newblock Sci. Rep. {\bf 5}, 12309 (2015).

\bibitem{Michaeli_PRL2012}
K.~Michaeli and A.~C. Potter and P.~A. Lee,
\newblock Phys. Rev. Lett. {\bf 108}, 117003 (2012).

\bibitem{Banerjee2013}
S.~Banerjee and O.~Erten and M.~Randeria,
\newblock Nat. Phys. {\bf 9}, 626 (2013).

\bibitem{Kelly_PRB2014}
N.~Ganguli and P.~J. Kelly,
\newblock Phys. Rev. Lett. {\bf 113}, 127201 (2014).

\bibitem{Pavlenko_PRB2009}
N.~Pavlenko,
\newblock Phys. Rev. B {\bf 80}, 075105 (2009).

\bibitem{Vanderbilt_PRB2009}
M.~Stengel and D.~Vanderbilt,
\newblock Phys. Rev. B {\bf 80}, 241103 (2009).

\bibitem{Park_PRL2013}
J.~Park, B.-G. Cho, K.~D. Kim and J.~Koo and H.~Jang and K.-T. Ko and J.-H.
  Park and K.-B. Lee and J.-Y. Kim and D.~R. Lee and C.~A. Burns and S.~S.~A.
  Seo and H.~N. Lee,
\newblock Phys. Rev. Lett. {\bf 110}, 017401 (2013).

\bibitem{Pavlenko_PRB2013}
N.~Pavlenko and T.~Kopp and J.~Mannhart,
\newblock Phys. Rev. B {\bf 88}, 201104 (2013).

\bibitem{Stemmer_PRX2012}
P.~Moetakef and J.~R. Williams and D.~G. Ouellette and A.~P. Kajdos and
  D.~Goldhaber-Gordon and S.~J. Allen and S.~Stemmer,
\newblock Phys. Rev. X {\bf 2}, 021014 (2012).

\bibitem{Lee_Nature2013}
J.-S. Lee and Y.~W. Xie and H.~K. Sato and C.~Bell and Y.~Hikita and H.~Y.
  Hwang and C.-C. Kao,
\newblock Nat. Mater. {\bf 12}, 703 (2013).

\bibitem{Bert_NPhys2011}
J.~A. Bert and B.~Kalisky and C.~Bell and M.~Kim and Y.~Hikita and H.~Y. Hwang
  and K.~A. Moler,
\newblock Nat. Phys. {\bf 7}, 767 (2011).

\bibitem{CNayak_PRB2013}
L.~Fidkowski and H.-C. Jiang and R.~M. Lutchyn and C.~Nayak,
\newblock Phys. Rev. B {\bf 87}, 014436 (2013).

\bibitem{Pavlenko_PRB2012_1}
N.~Pavlenko and T.~Kopp and E.~Y. Tsymbal and G.~A. Sawatzky and J.~Mannhart,
\newblock Phys. Rev. B {\bf 85}, 020407 (2012).

\bibitem{mohantaJPCM}
N.~Mohanta and A.~Taraphder,
\newblock J. Phys.: Condens. Matter {\bf 26}, 025705 (2014).

\bibitem{NM_VacancyJPCM2014}
N.~Mohanta and A.~Taraphder,
\newblock J. Phys.: Condens. Matter {\bf 26}, 215703 (2014).

\bibitem{Ariando2011}
Ariando, X.~Wang, G.~Baskaran, Z.~Q. Liu and J.~Huijben and J.~B. Yi and
  A.~Annadi and A.~R. Barman and A.~Rusydi and S.~Dhar and Y.~P. Feng and
  J.~Ding and H.~Hilgenkamp and T.~Venkatesan,
\newblock Nat. Commun. {\bf 2}, 188 (2011).

\bibitem{2014arXiv1411.3103K}
S.~{Kumar} and G.~{Nath Daptary} and P.~{Kumar} and A.~{Dogra} and N.~{Mohanta}
  and A.~{Taraphder} and R.~C. {Budhani} and A.~{Bid},
\newblock arxiv: 1411.3103 (2014) .

\bibitem{0295-5075-108-6-60001}
N.~Mohanta and A.~Taraphder,
\newblock EPL {\bf 108}, 60001 (2014).

\bibitem{Scheurer2015}
M.~S. Scheurer and J.~Schmalian,
\newblock Nat. Commun. {\bf 6}, 6005 (2015).

\bibitem{Shalom_PRL2010}
M.~Ben~Shalom and M.~Sachs and D.~Rakhmilevitch and A.~Palevski and Y.~Dagan,
\newblock Phys. Rev. Lett. {\bf 104}, 126802 (2010).

\bibitem{Held_PRB2013}
Z.~Zhong and A.~T\'oth and K.~Held,
\newblock Phys. Rev. B {\bf 87}, 161102 (2013).

\bibitem{Khalsa_PRB2013}
G.~Khalsa and B.~Lee and A.~H. MacDonald,
\newblock Phys. Rev. B {\bf 88}, 041302 (2013).

\bibitem{PhysRevB.92.174531}
N.~Mohanta and A.~Taraphder,
\newblock Phys. Rev. B {\bf 92}, 174531 (2015).

\bibitem{Nakamura_JPSJ2013}
Y.~Nakamura and Y.~Yanase,
\newblock J. Phys. Soc. Jpn. {\bf 82}, 083705 (2013).

\bibitem{Caprara_PRB2013}
S.~Caprara and J.~Biscaras and N.~Bergeal and D.~Bucheli and S.~Hurand and
  C.~Feuillet-Palma and A.~Rastogi and R.~C. Budhani and J.~Lesueur and
  M.~Grilli,
\newblock Phys. Rev. B {\bf 88}, 020504 (2013).

\bibitem{Imada_JPSJ}
Y.~Yamaji and T.~Misawa and M.~Imada,
\newblock J. Phys. Soc. Jpn. {\bf 75}, 094719 (2006).

\bibitem{Joshua_PNAS2013}
A.~Joshua and J.~Ruhman and S.~Pecker and E.~Altman and S.~Ilani,
\newblock Proc. Natl. Acad. Sci. USA {\bf 110}, 9633  (2013).

\bibitem{Joshua_NComm2012}
A.~Joshua and S.~Pecker and J.~Ruhman and E.~Altman and S.~Ilani,
\newblock Nat. Commun. {\bf 3}, 1129 (2012).

\bibitem{ncomm_nagaosa}
C.~Cancellieri and A.~S. Mishchenko and U.~Aschauer and A.~Filippetti and
  C.~Faber and O.~S. Barisic and V.~A. Rogalev and T.~Schmitt and N.~Nagaosa
  and V.~N. Strocov,
\newblock Nat. Commun. {\bf 7}, 10386 (2016).

\bibitem{imada}
M.~Hirayama and T.~Miyake and M.~Imada,
\newblock J. Phys. Soc. Jpn. {\bf 81}, 084708 (2012).

\bibitem{note1}
For example, in BaOsO$_{3}$/BaTiO$_{3}$, an atomic SOC $\sim 440$ meV leads to
  a gap of 570.4 meV, while the 19.3 meV SOC in the present case produces a 19
  meV gap.

\bibitem{Kotliar_PRL1998}
G.~P\'alsson and G.~Kotliar,
\newblock Phys. Rev. Lett. {\bf 80}, 4775 (1998).

\bibitem{Triscone_PRB2010}
I.~Pallecchi and M.~Codda and E.~Galleani~d'Agliano and D.~Marr\'e and A.~D.
  Caviglia and N.~Reyren and S.~Gariglio and J.-M. Triscone,
\newblock Phys. Rev. B {\bf 81}, 085414 (2010).

\bibitem{Valla_PRL2000}
T.~Valla and A.~V. Fedorov and P.~D. Johnson and J.~Xue and K.~E. Smith and
  F.~J. DiSalvo,
\newblock Phys. Rev. Lett. {\bf 85}, 4759 (2000).

\end{thebibliography}

\end{document}